# An Ontological Analysis of a Proposed Theory for Software Development

Diana Kirk ✉
*Independent Research Consultancy,*
*Auckland, New Zealand*
dianakirk@acm.org

Stephen G. MacDonell
*Software Engineering Laboratory (SERL),*
*Auckland University of Technology (AUT),*
*Auckland, New Zealand*
stephen.macdonell@aut.ac.nz

**Abstract**

*There is growing acknowledgement within the software engineering community that a theory of software development is needed to integrate the myriad methodologies that are currently popular, some of which are based on opposing perspectives. We have been developing such a theory for a number of years. In this paper, we overview our theory and report on a recent ontological analysis of the theory constructs. We suggest that, once fully developed, this theory, or one similar to it, may be applied to support situated software development, by providing an overarching model within which software initiatives might be categorised and understood. Such understanding would inevitably lead to greater predictability with respect to outcomes.*

**Keywords:** Software development, Software engineering, Theoretical model, Ontology, Software context

## 1. INTRODUCTION

The term *Software Engineering* was coined in 1968 at a conference whose aim was to discuss the need for the software development discipline to be more strongly based on theoretical and engineering principles [46]. The *Waterfall* model, a then-popular model used in manufacturing, was adopted as the standard app- roach for developing computer software. As time progressed, it became apparent that a strict implementation of this model was not appropriate for all software. A number of modifications, for example *Spiral* [8], and alternative models, for example *XP* [7], have emerged. The authors of the various models have different viewpoints on what kind of activity software development actually is. Earlier models view software development as an engineering activity and focus on control. More recent models adopt the viewpoint of 'software-as-a-service' and focus on effective communications. However, regardless of the huge variation in app- roach, until recently, the accepted wisdom by all methodology architects was that, in order to be fully effective, their approach must be followed exactly, with nothing added and nothing missed [12].

We have long understood from experiences in industry that this 'wisdom' is not based on what actually happens in the field, and have advocated with others the need to more deeply understand the process of developing computer software in order to support industry in its need to select practices in a flexible way, according to objectives and context [4,15,25,30,35,36]. This viewpoint has now become the accepted one [2,3,44,49,55]. The traditional viewpoint — that methodologies and practices should be adopted and used as prescribed — has thus been superseded by one of acceptance that tailoring is both necessary and unavoidable.

If tailoring is of the essence, we clearly must strive to fully understand the nature of the relationships between objectives, process and context. Only then will we be in a position to advise industry on which practices might be most suitable or to predict project outcomes based on context and practices implemented. The sole route to this kind of general understanding is through theory-building [16]. Without an understanding of the relationships between objectives, practices, context and outcomes, we can at best expose patterns based on data or observations. Such patterns represent *correlations* and correlations can not be used to predict in a general way. For consistent prediction, we must create frameworks based on causal relationships i.e. theoretical models. Indeed, Basili et al. remind us that, when carrying out controlled experiments, "... it is hard to know how to abstract important knowledge without a framework for relating the studies" [6].

The role of theory in software engineering (SE) has been investigated from a number of perspectives. Sjøberg et al. observed that there is very little focus on theories in software engineering and remind us of the key role played by theory-building if we wish to accumulate knowledge that may be used in a wide range of settings [51]. Hannay et al. conducted a review of the literature on experiments in SE and found that fewer than a third of studies applied theory to explain the cause-and-effect relationship(s) under investigation, and that a third of the theories applied were themselves proposed by the article authors [24]. Gregor examined the nature of theory in Information Systems (IS) and found multiple views on what constitutes a theory [17]. Stol and Fitzgerald argue that SE research does, in fact, exhibit traces of theory and suggest that the current focus on evidence based software engineering (EBSE) must be combined with a theory-focussed research approach to support explanation and prediction [52].

While there is general agreement within the research community that an increased focus on theory building would produce benefits, there remains uncertainty about how to proceed. Wand et al. remind us that "To employ conceptual modeling constructs effectively, their meanings have to be defined rigorously" and that achievement of this requires an articulation within the context of ontology [56]. The authors apply an ontological framework developed by Bunge [9] to analyse the relationship construct in entity-relationship modelling. A number of authors have used their approach to investigate various aspects of IS, for example, UML [47] and reference models [14]. A key aspect of the approach is the ability to confirm that a model is complete and without redundancy.

Our aim is to establish a conceptual model to describe a *software initiative*. By *software initiative*, we mean the software-related processes implemented to achieve specified outcomes. We suggest that the existence of such a model would support the software industry in the selection of appropriate practices, according to an organisation's specific objectives and contexts.

In this paper, we report on an ontological analysis of our proposed theory. We analysed the theoretical constructs with respect to the Unified Foundational Ontology (UFO) [21] with a view to identifying issues relating to meaning. The aim of our analysis was to gain a better understanding of the constructs that form the basis of our model. In Sect. 2 we discuss other efforts towards providing a theoretical foundation for software development. In Sect.3, we overview our proposed model prior to ontological analysis. In Sect.4, we present the results of our analysis based on the UFO and in Sect.5, we summarise the paper and discuss limitations and future work.

## 2. RELATED WORK

As far as we are aware, the only initiative to create a theory specifically for software is the SEMAT initiative, launched in 2009 by Ivar Jacobson, Bertrand Meyer and Richard Soley [27]. The approach proposes a *SEMAT* kernel comprising three parts. The first is a means of measuring project progress and health, the second categorises the activities required to effect progress, and the third defines the competencies required to effect the activities [26]. There are seven top-level 'alphas' - *Requirements, Software System, Work, Team, Way of Working, Opportunity* and *Stakeholders*. It is claimed that these concepts support determination of a project's health and facilitate selection of a suitable set of practices.

Although the authors of the SEMAT approach state that the initiative pro- motes a "non-prescriptive, value-based" philosophy that encourages selection of practices according to context, we suggest that the approach is, in fact, prescriptive in intent. Each of the SEMAT elements has a number of states "which may be used to measure progress and health" [28], and this implies that the health of every software project can be measured in a common way. We do not believe this to be the case, because of the vast number of possible objectives and contexts. In addition, the SEMAT model appears to offer no guidance on how to choose activities. Other than mentioning competencies required for activities, there is no link between activities and objectives or context. How do we know if an activity is suitable when the developers are in different countries or when the team comprises multiple cultures? The authors state that gaps and overlaps in activities can be easily identified, but it is not clear how this can be achieved in an objective way. We also note that the theory relates to a software project, a scope we have identified as being too narrow. For the above reasons, we have issue with the claim that a general theory is being developed. We cannot see how the approach will help researchers better understand practice limitations.

Weiringa, discussing the field of requirements engineering, reminds us of the dangers of confusing solution design with research [57]. We suggest that SEMAT, at this point, represents a design initiative rather than a theory building exercise.

## 3. THEORY OVERVIEW

Our first observation is that we must scope our model to include any software initiative i.e. the scope is larger than a software *project*. Our rationale is the need to consider in a holistic way the entire *software-in-a-system (SIAS)* i.e. the software product or service as part of a larger whole. This 'whole' will vary in time as the software is first created and then used. The larger system during creation will include the development organisation, test environments (possibly stand-alone) and the client. After deployment, the software will become part of a system comprising any of hardware, software, humans and processes. The job is to create, deliver and sustain healthy software systems over a lifetime of operation. The reason behind our viewpoint of a need for greater holism has its roots in the uncertainty that results from the growing complexity of software systems. This complexity makes it impossible to categorise in a simple way the environment during development and makes it difficult to anticipate all future conditions under which the software will run i.e. we can no longer assume a stable and bounded operating environment. Conditions such as technology change and inappropriate use will probably affect the in-situ efficacy of the software. Uncertainty also characterises delivery mechanisms — in the past, a project developed a software product and then delivered this to a known client base. More recently, the web-based mechanism of 'deliver a little, solicit immediate client feedback, and deliver the next increment' has become popular [53], leaving the concept of a 'project-to-develop-then-deliver' no longer tenable.

As early as 1987, Basili and Rombach believed that "Sound tailoring requires the ability to characterise ... goals ..., the environment ..., and the effect of methods and tools on achieving these goals in a particular environment" [5]. Our efforts thus far have focused on clarifying what this might mean when considering flexibility in software initiatives. We thus begin with a consideration of *objectives*, *process* and *context*. In the following sections, we overview our efforts and comment on understandings achieved.

## 3.1 Objectives

We first observe the need to consider more than one objective for an initiative [33,36]. The reason is that a focus on a single outcome may lead to the identification of a local maximum and a possible sub-optimisation of the whole system [40,42,43]. We next observe that there are many possible objectives, including the common product-related ones of minimisation of cost and maximisation of quality, but also including the people-related ones such as 'increase developer subject-area knowledge', 'retain developers' or 'keep a specific customer happy'. For some of these, an associated numerical value will change throughout the initiative, for example, when spending increases throughout a project. For others, the goal is less definitive and a more fuzzy measure may be appropriate, for example, developer satisfaction levels may be described as 'Low', 'Medium' and 'High' [31]. A third observation is that most software initiatives are characterised by uncertainty [1,48] and this means that values cannot be represented in a deterministic way. A probabilistic distribution is a more suitable measure for some kinds of objective [11,39,50]. Our proposal is that a set of objectives may be modelled by a set of *values*, the type of value for each depending upon the nature of the objective. An implementation would involve representing an objective by a name, a description and a desired value.

Monitoring of progress, i.e. 'current state', will involve a consideration of the current value for each objective in the set. This takes into account the fact that state values are generally not 'empty' at commencement. For example, developers will have a certain 'satisfaction' level at commencement; in a product line situation, there is an existing code base that is characterised by a level of quality. While we believe this to be a sound approach, we have as yet taken this investigation no further i.e. the structure of *objectives* is an open research question.

## 3.2 Process

We view a process as a set of *practices*. Each Practice has the effect of moving the initiative closer to (or further away from) its *Objectives*. We observe that, in addition to the accepted practices such as 'design inspection', our definition includes anything that has an effect. For example, in a small startup organisation, an informal, unplanned practice such as 'chat over lunch' may be crucial for supporting the developer's understanding of what is to be built and is thus an acceptable 'practice' in our model. According to our model, a desired practice is one that successfully moves the initiative in the right direction. A 'lean' process is one in which every practice is effective.

The space of all possible practices is huge, and includes practices for identifying the target audience for a proposed product and understanding what the product should do, practices for designing, implementing and delivering the product, and practices for supporting product use in the target environment(s). We clearly need to introduce some structure, but found that the standard reference models did not suit when attempting to elicit information from individuals in smaller, less formal organisations [37]. It was clear that, if we wanted to capture practices-as-implemented-in-the-real-world, some new perspective was required.

Table 1. Categories for practices.

| Define | Roadmap |
|---|---|
| | Scope |
| Make | Design |
| | Implement |
| | Integrate |
| Deliver | Release |
| | Support |

Our approach considers what organisations *need to achieve* at a high level when involved in a software initiative. Our top-level functional categories are

– Define what is to be made.
– Make it.
– Deliver it.

We extended these categories to be what we believe are the main sub- categories for software. These are shown in Table 1.

Because we have structured based on *function*, the categorisation will support any practice deemed to be relevant for meeting objectives. Informal meetings in the lunch room that help developers understand scope clearly fit into the 'scope' category. We hypothesise that the proposed categorisation addresses all software practices. A precondition that an objective be met is that each category must contain one or more effective practices. To illustrate, for a 'Quality' objective, including quality considerations during product design, implementation and integration will fail to achieve the objective if quality expectations are not included during scoping. The identification of gaps is straightforward.

The categorisation above does *not* imply any ordering of practices. Such ordering would exist at a higher level and might be used to describe strategies of iteration and incremental delivery. We also submit that the categorisation is 'paradigm-agnostic' — whether an initiative is run in a traditional or agile way, the basic functions of defining, making and delivering the product must be carried out. Of note is the fact that the sub-categories 'Roadmap' and 'Support' lie outside of a traditional development project and sub-categories 'Scope' and 'Release' relate to practices that span development organisation and client.

Our work on the *practices* aspect of the proposed theory is embryonic. Testing thus far is limited to a single, exploratory study in which we captured practices in three New Zealand software organisations [37,38]. We did expose some interesting areas for further study. For example, all participating organisations reported a dependence on practices that involved having to actively search for information, possibly implying some inefficiency as individuals must spend time. However, the study was exploratory in nature and this is an open research area.

### 3.3 Context

This represents the most challenging aspect of any theory for software development. There have been many attempts to relate project outcomes to specific contextual factors, for example, [2, 10, 41, 53]. Our main critique of existing approaches is that they remain factors based [32]. We suggest that such an app- roach is misguided because

- there are simply too many possible factors to take into consideration, and this number will increase as new paradigms for software are introduced.
- it is unlikely that any two projects will be exactly the same and so under- standing key factors for some is unlikely to be of use in a general way.

We believe that it is crucial that we develop an operationalisation of *context* that will be relevant for all software initiatives i.e. takes into account the situated nature of a software product throughout its lifetime. It seems clear that the required model must comprise a number of orthogonal dimensions in order that an initiative can be plotted as a point in the dimensional space. In order to remove the 'factors-based' aspect, it is necessary to also find suitable abstractions for each dimension, abstractions that support a straight-forward identification of value for a given initiative. For example, rather than define a number of factors such as 'developer experience', 'developer subject area knowledge', we must abstract in such a way as to render it irrelevant if we have missed a factor out (for example, 'developer commitment').

Our first efforts at modelling this space involved a consideration of the dimensions *Who, Where, What, When, How and Why* [32]. These dimensions have been applied by others to ensure orthogonality [13,58]. Of course, the usefulness of the abstraction depends upon the choices about what these dimensions mean. Our assigned meanings are [34]:

**Who:** associates with peoples' *ability to perform*. Personal characteristics, culture and group structure are relevant, as these affect levels of understanding and conceptual sharing.

**Where:** associates with peoples' *availability*. The degrees of temporal and physical separation are relevant.

**What:** associates with *product* characteristics. Standards expectations, product interfaces and achieved quality are relevant.

**When:** associates with *product life cycle*. Examples are in-development, recently deployed, near retirement [45].

**How:** associates with *engagement expectations*. Client and developer expectations for the mechanisms for product specification and delivery will affect which practices are most appropriate.

**Why:** associates with *establishing objectives*.

We carried out some 'proof-of-concept' studies on this model, each involving categorising contextual factors from a small number of studies from the software engineering literature [29,34]. These studies caused us to refine our understanding of context in the following ways. In the first instance, it became clear that the dimension *why* addresses *objectives* and is thus not part of a model for *context*. We also found that many contextual factors mentioned in the literature are vague or ambiguous, and so must be clarified prior to categorising. For example, when considering the commonly-mentioned factor 'Company size', we suggest that it is not size itself that affects practice selection and/or outcomes, but rather what this means in terms of culture and physical and temporal separation. We labelled such factors as 'Secondary'. Some factors, such as 'requirements uncertainty', may be the result of one of a number of possible scenarios. For example, perhaps the client is not clear about what the product exactly is; perhaps (s)he is simply weak on decision-making; perhaps (s)he is unable to state what is wanted because of client-internal processes i.e. (s)he is waiting for a decision from someone else. We cannot know which practices will be effective until we understand which meaning is relevant. A practice of 'regular client meetings with prototypes' will not help if the client is waiting for someone else. We labelled this kind of factor as 'ambiguous'. Finally, we noticed that some factors were more 'high level' in that they could be more usefully viewed as affecting strategy. For example, 'lack of funds' would likely force some consideration about strategy, and the resulting decision may, in turn, affect objectives and/or context. It might be decided that the project should be abandoned, that some developers should be removed, or that a minimal product only should be delivered. We recognise these factors as *Strategic factors* and remove from our discussion of *context*.

Our current status is that, with our refined definition of *context*, we are well into a study to categorise contextual factors from the literature into our dimensional model. Thus far, we have met no obstacles. However, we have as yet included only literature from the *software engineering* domain and have constrained the study to the *development project*. Although we are optimistic, there clearly is much scope for research in this area, research that must be carried out before we can be confident in our proposed structure for *context*.

## 4. ONTOLOGICAL CONSIDERATIONS

As a preliminary to investigating the efficacy of our model in the field, we sought to more deeply understand its constructs from a *meaning perspective*. In order to achieve this, we turned to the field of *ontology*.

Ontology is primarily a long-established branch of philosophy that deals with the nature and structure of reality [19]. With the more recent growing interest in the semantic web, the term has been more generally used to represent a single domain in some knowledge representation language, for example, RDF and OWL [20]. However, the data modelling community has continued to apply the term in its more philosophical sense i.e. as a "philosophically well-founded domain-independent system of formal categories that can be used to articulate domain-specific models of reality" [20]. The claim is that, unless a domain model is based on the meaning of the domain concepts, the model will be flawed and open to mis-interpretation. One implication is that the use of logics for capturing systems is insufficient because logics do not address meaning. When discussing classifications of modelling primitives in knowledge representation, Guarino suggests that an ontological level for modelling is required to ensure meaning is adequately represented [18].

Guizzardi and Wagner suggest that an ontologically sound model is characterised by *truthfulness to reality (domain appropriateness)* and *conceptual clarity (comprehensibility appropriateness)* [23]. Without these constraints, a model may have several interpretations and cause misunderstandings in communication. These properties can be guaranteed if an isomorphism (one-to-one map-ping) exists between abstraction and domain models. Specifically, an abstraction must exhibit:

**Soundness** — each modelling construct maps to a domain construct.

**Completeness** — each domain construct is represented by a modelling construct.

**Lucidity** — each modelling concept represents at most one domain concept.

**Laconicity** — each domain construct is represented by at most one modelling construct.

Foundational ontologies are "theoretically well-founded domain-independent systems of categories that have been successfully used to improve the quality of conceptual ... models" [23]. They thus enable the description of general concepts to be used in conceptual modelling. To our knowledge, there are two main foundational ontologies that have been adopted by the Information Systems community. Each of these has been applied to expose issues in current modelling methods. The *Bunge-Wand-Weber* model is an ontology based on the work of the philosopher Bunge [9] and adapted by Wand and Weber for use in information systems [56]. The approach has been applied to expose issues with the relationship construct in entity-relationship modelling [56], and to analyse shortcomings in UML [47] and reference models [14]. The *Unified Foundational Ontology (UFO)* represents a "synthesis of a selection of foundational ontologies" with the goal of creating a foundational ontology that is "tailored towards applications in conceptual modelling" [21]. As the UFO appears to have been more fully developed, with greater consideration for process-related entities, we have chosen to use this as a basis for our analysis.

In the next section, we overview the UFO, including only those aspects that we believe are relevant to this research. In the following section, we categorise the components of our proposed theory in terms of the UFO.

### 4.1 Unified Foundational Ontology (UFO)

The Unified Foundational Ontology (UFO) aims to define a range of domain-independent ontological categories to be used as a basis and elaborated for specific domains. There are three layered sets [22]:

**UFO-A:** defines the core set and addresses *endurants* i.e. entities that do not depend upon time for identification, for example, a person or a document.

**UFO-B:** extends UFO-A to include *perdurants* i.e. entities that are critically time-dependent, for example, events and processes.

**UFO-C:** extends UFO-B to include terms relating to intentional and social things.

Endurants are changed by perdurants. For example, a document may be altered by the enactment of an editing procedure.

Some relevant terms are introduced below [23]. A fundamental distinction is made between the categories of Particular (Individual) and Universal (Type).

**Particulars (Individuals)** are entities that possess a unique identity. Examples are 'person', 'apple'.

**Universals (Types)** are patterns of features which can be realised in a number of different particulars. Examples are 'Person', 'Apple'.

**Substantials** are existentially independent particulars, for example 'apple'.

**Moments** are particulars that can exist only in other particulars and are thus existentially dependent upon them. These relate to the *properties* or *qualities* of a substantial. Examples, are a colour, a symptom.

**Intrinsic Moments** are dependent upon a single individual, for example, the *colour* of a rose.

**Relators (Relational Moment)** are dependent upon multiple individuals, for example, an *employment*. A relator can be viewed as the sum of the moments (properties) acquired by the participating individuals.

**Substance Universals** include, for example, 'Apple', 'Person'.

**Moment Universals** include, for example, 'Colour', 'Headache'.

**Relations** are universals that glue together other entities, for example, 'Employment'.

**Material Relations** have a material structure, for example, 'being employed by'.

**Formal Relations** hold between two or more entities directly, for example, 'part-of'.

**Quality Structure** provides the means for representing intrinsic moments, for example, a one-dimensional structure for representing 'height'.

**Quale** is a point in a quality structure.

**Conceptual Space** is a collection of quality structures, for example, a 'Colour' conceptual space may comprise structures for 'red', 'green' and 'blue'.

**Mode** is an intrinsic moment that can be conceptualised in terms of multiple separable quality dimensions, for example 'beliefs', 'symptoms', skills'.

At first glance, it would appear that the UFO-C would also be relevant, as we are aiming to model a software development initiative and this inherently includes teams, goals, etc. However, our aim is not to fully describe a software initiative, but is rather to find an abstraction that will support selection of suit- able practices. As we consider all contextual aspects as simply affecting practice efficacy, we have no need of the social aspects covered in UFO-C. In the next section, we report the results of considering our proposed abstraction in the light of the UFO-A and UFO-B.

### 4.2 Application of UFO to Proposed Theory

The concepts identified above as being relevant for our abstraction are *Software Initiative, Objectives, Context* and set of *Practice*.

In keeping with the convention of naming particulars and universals above, we use first letter uppercase for naming universal concepts and first letter lowercase for naming particulars.

We categorise Software Initiative as a substance universal i.e. a type, and software initiative as a substantial particular i.e. an instantiation.

When considering 'Objectives', we notice that objectives are closely bound to a software initiative i.e. are existentially dependent upon the initiative. We also observe that a software initiative has a current status, which changes as the initiative progresses and which has the same structure as objectives. It would seem reasonable to categorise both *Objectives* and *Current Status* as properties of Software Initiative i.e. as moment universals. As these constructs comprise multiple separable quality dimensions, for example, 'Time to market', 'Developer satisfaction', we represent as a *mode*.

We now consider Practice. We view as a substantial i.e. uniquely identifiable entity, with properties such as 'name', 'description' and 'category' (see Table 1). Torres reminds us that "sometimes we become so enamored with our favorite tools and techniques that we lose sight that they ... have a nominal operating range" [54]. We recognise that a practice is inherently associated with an operating specification, for example, 'client must be available' or 'effective change management for a fast-changing product but not for a stable one'. These constraints would be modelled as intrinsic moments of Practice. We observe that the operating specification relates directly to context, and revisit below.

On considering Context, our first analysis results in viewing as a substantial, with moments (properties) such as 'name' and 'description'. We have suggested in Sect.3.3 that context is 'made up of' a number of parts i.e. a who-part, a where-part, etc. Each of these parts would be represented as a substantial. A material relation 'implement practice within a context and software initiative', exemplified by a relator entity, let's say 'practice implementation', would then bind a specific software initiative, practice and context.

However, we noticed in the previous paragraph that a practice's operating specification is closely related to the concept of *context* in that an operating specification will comprise the same kinds of values as are found in context. If we view Operating Specification as an intrinsic moment (property) of Practice, we cannot easily relate this to Context-as-a-Substantial. We are applying different kinds of modelling construct to model domain constructs that have the same basis. This would compromise the conceptual clarity of our model.

Another viewpoint is to consider that Context might be better viewed as an intrinsic moment of Software Initiative i.e. is existentially dependent upon Software Initiative. However, the moment comes into being only when a practice is carried out within a software initiative i.e. it is externally dependent upon the Relation 'implement practice within a software initiative' and the associated relator (relational moment) 'practice implementation'. To illustrate, the developer characteristics and lifecycle stage associated with a software initiative become relevant only in relation to a practice implementation. For example, a developer may be extremely skilled in playing chess and in designing software systems, but neither of these is relevant for a task of contract negotiation with a potential client.

We now have both context and operating specification viewed as intrinsic properties of substantials. As Context comprises several separable dimensions (who, when, what, where and how), we model as a mode, each dimension of which is a quality structure. This supports analysis of whether a practice is being implemented within its operating constraints.

From UFO-B, which extends UFO-A, we can consider the perdurant 'practice implementation' as an event that causes a change of state in its participating substantials. In our abstraction, change is effected to both current status and context (both intrinsic moments of software initiative). For example, current status may have become closer to some objectives and developer experience may have increased.

In summary, our abstraction is:

**Software Initiative** is a substantial universal with moment universals Objectives and Current Status. The moments are described as a *Software Status* mode. For a specific software initiative, software status comprises the quality dimensions appropriate for the initiative (for example, cost) and the elements of each as quales.

**Practice** is a substantial universal with moment universals Category and Operating Constraints. The latter has a structure similar to that of *Context*.

**Practice Implementation** is a relational moment (relator), relating Software Initiative and Practice and described as 'a practice is implemented within a software initiative'. As a relator, Practice Implementation is characterised by the properties of the participating substantials that become relevant as a result of participation in the relationship. The relevant property is Context, a moment of software initiative. Context is a mode, with dimensions modelled as quality structures.

### 4.3 Discussion

We submit that the above abstraction complies with the requirements that a model be characterised by *domain appropriateness and comprehensibility appropriateness* [23]. We believe the exercise has been beneficial in that the need for separation of 'practice' and 'practice implementation' has been exposed. We have had to think more clearly about the roles of objectives and current status. We have exposed a need for the concept 'operating specification' and have considered its link with context.
Although the UFO extension, UFO-C, would appear to be relevant, in that it includes ideas such as 'objective' and 'intentions', we submit that, because we are not trying to describe a software development process, the concepts are not necessary for our purpose. We wish to more deeply understand how to sup- port practice selection and at this point believe that the influence of human attributes can be viewed in the same way as, for example, lifecycle stage for this purpose. Each of these affects which practices are (contra-)indicated for the software initiative.

An observation from the above analysis, is that the relational moment *context* is dependent upon the participation of a software initiative in a practice implementation relationship. As noted above, a developer's experience and location have meaning only in relation to the task being carried out. The implication is that discussing a software initiative's context in general has no meaning. This is potentially problematic from a practical perspective in that we cannot ascertain practice implementation efficacy without understanding which contextual elements are relevant and we cannot know which contextual elements are relevant without considering a practice implementation. We believe the key lies in how the dimensions of context are abstracted, and this is clearly an area for further research.

Although our approach has shown benefits, we are clear that this does not mean our proposed theory is correct or useful i.e. if our abstraction does not model a software initiative in a way that supports practice selection, its adherence to ontological requirements is irrelevant. Our goal is rather to expose potential issues with theory structure before embarking on a more extensive and situated evaluation.

## 5. SUMMARY

We have proposed an embryonic theory for software development with the aim of more deeply understanding which practices might be indicated and contra- indicated within specific contexts. One characteristic of a successful theory is that opposing viewpoints can be seen as facets of a larger whole. Our conceptualisation must address the traditional versus agile dichotomy. It is not difficult to see that, as a situated practice is defined simply as a transformation of current state, many different practices will exist that implement the same kind of transformation. For example, in relation to Table 1, the practices 'Formally document requirements' and XP's 'Planning Game' both result in an increased understanding of what is to be built. However, if we consider the objectives 'Reliability' and 'Increase developer subject area knowledge', it is likely that the formal documentation approach will address the former (as non-functional requirements are an inherent part of formal requirements documents) while 'Planning Game' will not. The XP approach may require additional practices to address quality expectations.

Before embarking on a more exhaustive consideration of the efficacy of our theory, we deemed it necessary to gain confidence in the proposed constructs from a 'meaning' perspective and have completed a preliminary ontological analysis. When mapping the theoretical constructs to constructs in the UFO (Unified Foundational Ontology), we identified Software Initiative and Practice as basic entities, understood that a relator entity 'Practice Implementation' is required and established that Objectives and Context can be represented as multi-dimensional modes. We believe our analysis has been of benefit in that we now have a mechanism for associating a software initiative's objectives and current state and a practice's operating constraints and implementation context.

In summary, we have presented an overview of the theoretical approach we have been pursuing for several years and a report of our recent ontological analysis. Our work is in-embryo. Our contribution is that we are making an honest attempt to tackle an extremely difficult problem, and have set a new direction for exploration. We believe we have made pockets of progress in some areas. Our hope is that the efforts we have made thus far will be used as a starting point for other researchers. We submit that, if the community does not take the theory-building initiative seriously, we are doomed to endless cycles of 'new' process paradigms and architectures, each of which has some merit and many shortfalls.